\documentclass[a4paper,11pt]{article}
\usepackage{jheppub} 
\usepackage{color}
\usepackage{subcaption}
\usepackage[normalem]{ulem}
\usepackage{tikz}
\usepackage{mathdots}
\usepackage{comment}
\usetikzlibrary{arrows.meta,positioning}
\usepackage{enumitem}
\usetikzlibrary{calc}

\newcommand{\comm}[1]{} 

\def\({\left(}
\def\){\right)}
\def\[{\left[}
\def\]{\right]}

\def\rma{{\mathrm{a}}}
\def\rmb{{\mathrm{b}}}

\def\One{{\hbox{ 1\kern-.8mm l}}}

\def\barray{\begin{array}}
\def\earray{\end{array}}
\def\be{\begin{equation}}
\def\ee{\end{equation}}
\def\bea{\begin{eqnarray}}
\def\eea{\end{eqnarray}}
\def\bal{\begin{align}}
\def\eal{\end{align}}
\def\nn{\nonumber}

\def\-{\,-\,}
\def\={\,=\,}
\def\+{\,+\,}
\def\equi{\,\equiv\,}

\numberwithin{equation}{section} 

\definecolor{cardinal}{rgb}{0.6,0,0}
\definecolor{darkgreen}{rgb}{0,0.4,0}
\definecolor{golden}{rgb}{0.92, 0.7, 0}
\definecolor{midnight}{rgb}{0, 0, 0.5}
\definecolor{darkblue}{rgb}{0, 0, 0.7}
\definecolor{purple}{rgb}{0.5, 0, 0.5}

\definecolor{amaranthred}{rgb}{0.83,0.13,0.18}
\definecolor{amazon}{rgb}{0.23,0.48,0.34}
\definecolor{bdazzledblue}{rgb}{0.18,0.35,0.58}
\definecolor{absolutezero}{rgb}{0.0,0.28,0.73}
\definecolor{bitterlemon}{rgb}{0.79,0.88,0.05}
\definecolor{byzantine}{rgb}{0.74,0.2,0.64}
\definecolor{turquoise}{rgb}{0.19, 0.84, 0.78}
\definecolor{burgundy}{rgb}{0.5, 0.0, 0.13}

\def\IR{\mathbb{R}}

\def\cM{{\cal M}}

\def\cP{{\cal P}}
\def\cO{{\cal O}}

\def\cW{{\cal W}}
\def\cX{{\cal X}}

\def\cZ{{\cal Z}}

\title{Generating Rotation in a Snap}

\author[a,b]{Soumangsu Chakraborty,} 
\author[b]{Pierre Heidmann,}
\author[b]{Gela Patashuri}

\affiliation[a]{Department of Physics\\	
Indian Institute of Technology Kharagpur, Kharagpur 721302, India}
\affiliation[b]{Department of Physics,
Center for Cosmology and AstroParticle Physics (CCAPP)\\
The Ohio State University,
191 W Woodruff Ave, Columbus, OH 43210, USA}

\emailAdd{soumangsuchakraborty@gmail.com, heidmann.5@osu.edu, patashuri.1@osu.edu}

\abstract{
We build a technique to generate rotation from arbitrary static solutions that asymptote to four- or five-dimensional Minkowski spacetime based on Clément's approach \cite{Clement:1997tx}. The method is purely algebraic and does not require solving Einstein equations. It proceeds by transforming the static solution to AdS$\times$S asymptotics, performing a coordinate shift to a uniformly rotating frame, and then transforming the solution back to asymptotically flat spacetime. We implement this construction in five-dimensional minimal supergravity, although it applies more broadly to any framework admitting AdS$\times$S geometries and relevant sigma-model transformations. As a first application, we recover simply the Kerr and Myers-Perry black holes directly from Schwarzschild black holes. We then apply the method to the linear class of static Weyl solutions and obtain the first linear ansatz describing an arbitrary number of non-extremal rotating and charged sources. This approach provides a systematic and simple route to constructing non-extremal rotating geometries in four and five dimensions.
}

\begin{document}
\maketitle
\flushbottom

\newpage
\section{Introduction}
\label{sec:intro}

As illustrated by the fifty years that elapsed between the Schwarzschild solution and the introduction of rotation in the Kerr black hole \cite{Kerr:1963ud}, constructing rotating geometries from Einstein’s equations is notoriously challenging. Historically, such solutions have been obtained by solving the full nonlinear Einstein equations directly. This typically requires exploiting integrable structures that emerge upon dimensional reduction to three dimensions, where the system can be recast as a \textit{nonlinear sigma model}. A prominent example is the Ernst formalism for four-dimensional Einstein-Maxwell theory \cite{Ernst:1967wx,Ernst:1967by}. Within this framework, powerful techniques such as inverse scattering and monodromy methods have been developed \cite{Belinsky:1971nt,Belinsky:1979mh,Belinski:2001ph,Manko_1993}, ultimately leading to the construction of intricate solutions describing multiple non-extremal black holes on a line \cite{Ruiz:1995uh,NoraBreton1998}. However, these approaches depend significantly on the four-dimensional theory and lack a universal, systematic prescription. 

Significant efforts have therefore been devoted to developing simpler, more systematic procedures to generate rotation through \textit{algebraic transformations} that avoid solving differential equations. A well-known example is the Newman-Janis algorithm \cite{Newman:1965tw}, which produces the Kerr metric from Schwarzschild via a complex coordinate transformation. However, this procedure is best viewed as a heuristic trick rather than a systematic method, as it does not readily generalize beyond a narrow class of solutions \cite{Azreg-Ainou:2014pra}.

More generally, nonlinear sigma models possess hidden symmetries that allow for solution-generating transformations. These transformations, such as the Harrison-Ehlers transformations in the Ernst formalism \cite{harrison_new_1968,Geroch:1970nt,PhysRevLett.41.1197,Harrison:1980fr,Alekseev:2020bqu}, provide an efficient way to generate new solutions from a given seed. For instance, it is straightforward to ``charge up'' a vacuum solution using such transformations. However, because static solutions are typically made of monopole sources and these transformations merely reshuffle existing degrees of freedom, they cannot directly generate the dipole structure required for rotation. 

A notable way out was identified by Clément \cite{Clement:1997tx,Clement:1998nk,Clement:1999bv}, who demonstrated that rotation can nevertheless be generated from static solutions using Harrison-Ehlers transformations in four dimensions. The key idea is to transpose the seed solution with such a transformation to an asymptotically AdS$_2\times$S$^2$ geometry, perform a coordinate shift to a uniformly rotating frame, and then transform the solution back to asymptotically flat spacetime. This procedure effectively induces angular momentum through a sequence of algebraic transformations. However, its applicability is limited by its reliance on the four-dimensional Ernst formalism and, more generally, on four-dimensional gravity, where uniqueness theorems severely constrain the space of physically meaningful solutions.

In this work, we generalize this approach to spacetimes of dimension higher than four, which may be additional compact directions, and to a broader class of theories that admit a three-dimensional nonlinear sigma-model structure and allow for transformations between asymptotically flat and AdS$\times$S geometries. We focus on minimal five-dimensional supergravity, although our method extends straightforwardly to the STU model and to eleven-dimensional supergravity on T$^6$, as well as to other dual supergravity frameworks. This theory exhibit a rich sigma-model structure with a $G_{2(2)}$ symmetry \cite{cremmer_1989,Chamseddine:1980mpx,Bouchareb:2007ax,Clement:2007qy}.\footnote{We thank Gerard Clément for pointing out that the method in five-dimensional minimal supergravity was already sketched in Section 4.3.2 of \cite{Clement:2008qx}.}

\begin{figure}[t]
\centering
\begin{tikzpicture}[
    >=Stealth,
    node distance=3.4cm and 4.6cm,
    box/.style={
        draw,
        rounded corners=3pt,
        thick,
        fill=gray!15,
        text width=4.1cm,
        minimum height=1.6cm,
        align=center,
        inner sep=6pt
    },
    arrow/.style={
        ->,
        line width=1.3pt,
        draw=black!80
    }
]

\node[box] (box1) {Static geometry\\ in flat space};

\node[box, right=of box1]
(box2) {Static geometry\\ in AdS$\times$S space};

\node[box, below=2cm of box2]
(box3) {AdS$\times$S geometry\\ in a rotating frame};

\node[box, left=of box3]
(box4) {Rotating geometry\\ in flat space};

\draw[arrow]
(box1) -- node[above,align=center]
{Sigma-model\\transformation} (box2);

\draw[arrow]
(box2) -- node[right,align=center]
{Coordinate\\shift} (box3);

\draw[arrow]
(box3) -- node[above,align=center]
{Sigma-model\\transformation} (box4);

\end{tikzpicture}

\caption{Schematic diagram illustrating the solution-generating procedure used to obtain rotating solutions from static geometries in $\mathbb{R}^{1,3}\times$S$^1$ or $\mathbb{R}^{1,4}$.}

\label{pic:genmap4d}
\end{figure}
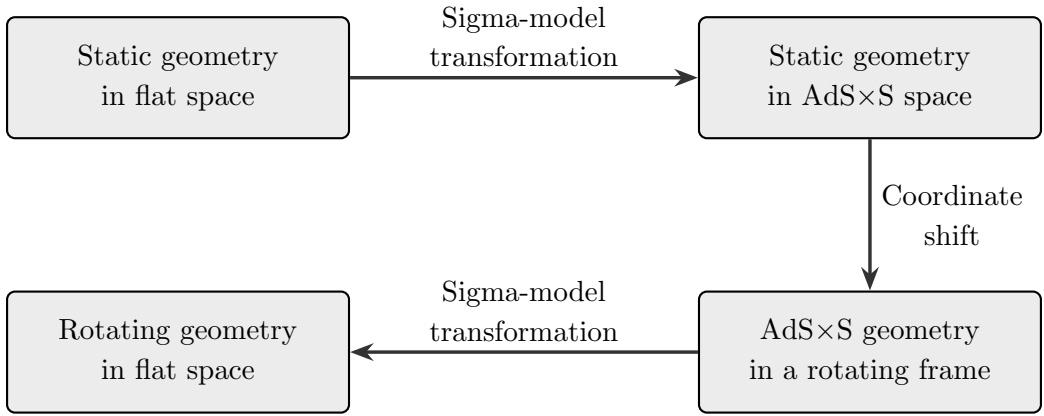

The essential steps follow Clément's approach, are summarized in Fig.\ref{pic:genmap4d}, and, importantly, involve only algebraic manipulations, without the need to solve differential equations. Our approach allows one to generate rotation for arbitrary static solutions that are asymptotic to $\mathbb{R}^{1,3}\times$S$^1$ or $\mathbb{R}^{1,4}$. We first identify the sigma-model transformations that map these asymptotics to AdS$_3\times$S$^2$ or AdS$_2\times$S$^3$. We then provide an explicit and systematic recipe to generate rotation using these transformations together with a coordinate shift.

As a first application, we show that our technique allows one to generate the Kerr black hole from the Schwarzschild metric through a sequence of simple steps, in agreement with Clément’s four-dimensional construction. Importantly, it extends successfully to higher dimensions, yielding in particular the five-dimensional Myers-Perry black hole \cite{Myers:1986un,Myers:2011yc} from the Schwarzschild-Tangherlini metric.

To further demonstrate the power and generality of our method, we apply it to Weyl solutions, which describe the most general static, axisymmetric configurations in four dimensions in terms of a linear Laplace equation \cite{Weyl:book,Israel1964,Emparan:2001wk}. By embedding them in five dimensions with a trivial S$^1$, our construction yields the first linear ansatz describing an arbitrary number of non-extremal, rotating, and charged sources aligned along a line. Previously, such linear ansatz were restricted to extremal (BPS) configurations, where electromagnetic repulsion balances gravitational attraction and allows for linear superposition.

While the rotation-generating procedure is systematic, it is important to emphasize two limitations. First, the intermediate mapping to AdS$\times$S geometries generically turns on gauge fields and charges. Although these can often be adjusted using standard sigma-model transformations to recover neutral solutions at the end, there is no guarantee that the final configuration is purely vacuum, even when starting from a vacuum seed. Residual dipole charges may persist. Second, although the number of sources is preserved, their local geometric structure may change under the transformations and coordinate shift. In particular, loci corresponding to smooth degeneracies of spacelike circles in the seed solution may not retain the same topology after the transformation.

Finally, this rotation-generating technique provides, in particular, a new and simple tool for constructing smooth, horizonless geometries with angular momentum. Over the past few years, significant progress has been made in constructing static non-extremal solitonic solutions, such as topological stars \cite{Bah:2020ogh,Bah:2020pdz,Bah:2021owp,Bah:2021rki}, $\cW$-solitons \cite{Chakraborty:2025ger,Dima:2025tjz}, and Schwarzschild solitons \cite{Bah:2022yji,Bah:2023ows}, which capture aspects of black hole microstructure beyond supersymmetry. However, extending these constructions to include rotation has remained a major challenge. Recent advances based on sigma-model techniques have begun to address this issue \cite{Chakraborty:2025ger,Heidmann:2025pbb,Bianchi:2025uis}. The method developed here opens the door to systematically adding rotation to this entire class of static solutions, thereby significantly expanding the landscape of physically relevant horizonless geometries.

The paper is organized as follows. In Section \ref{sec:SigmaModel}, we briefly review the nonlinear sigma model arising from five-dimensional supergravity and identify the relevant transformations. In Section \ref{sec:KerrGen}, we present our rotation-generating technique, which algebraically induces angular momentum in asymptotically $\mathbb{R}^{1,3}\times$S$^1$ geometries. As applications, we show how to obtain the Kerr solution from Schwarzschild and derive a linear ansatz describing rotating and charged configurations from the static Weyl solutions. In Section \ref{sec:MPGen}, we develop a similar procedure for asymptotically $\mathbb{R}^{1,4}$ geometries and demonstrate how the Myers-Perry black hole simply arises from the Schwarzschild-Tangherlini solution. We conclude with a discussion and directions for future work in Section \ref{sec:Discussion}.

\section{Sigma model in five-dimensional supergravity}
\label{sec:SigmaModel}

In this section, we review the formalism required to generate rotations via sigma model and coordinate transformations. In principle, any theory satisfying the following criteria would be suitable:
\begin{itemize}[topsep=0pt,itemsep=0pt,parsep=0pt,partopsep=0pt]
\item It admits a sigma model with target-space symmetries upon reduction to three dimensions.
\item It contains solutions of the type $\mathrm{AdS}\times$S, typically requiring at least one gauge field.
\end{itemize}
\vspace{0.2cm}

We will work with five-dimensional $\mathcal{N}=2$ minimal supergravity, which arises as a consistent truncation of STU supergravity, and admits a sigma model with $G_{2(2)}$ target-space symmetry \cite{cremmer_1989,Chamseddine:1980mpx,Bouchareb:2007ax,Clement:2007qy}. The action is
\begin{equation}
    \mathcal{S}_5=\frac{1}{16 \pi G_5} \int\left(R \star_5 1-\frac{3}{2} F \wedge \star_5 F-F \wedge F \wedge A\right),\qquad F=d A.
\end{equation}
where $A$ is a $\mathrm{U}(1)$ gauge field.

\subsection{Brief review}
\label{eq:ReviewSM}

We consider solutions admitting two commuting Killing symmetries, one timelike and one spacelike, parametrized by the coordinates $t$ and $\psi$, respectively. The five-dimensional fields are decomposed as
\begin{equation}
    \begin{aligned}
        d s_5^2 & =-\frac{1}{Z^{2}}\left(d t+\mu\left(d \psi+\omega_\psi\right)+\omega_t\right)^2+Z\left[\frac{1}{Z_0}\left(d \psi+\omega_\psi\right)^2+Z_0 d s_3^2\right]~, \\
    A & =A_t\left(d t+\mu\left(d \psi+\omega_\psi\right)+\omega_t\right)+A_\psi\left(d \psi+\omega_\psi\right)+a~,
    \end{aligned}
\label{eq:Ansatz}
\end{equation}
where the fields are described by five scalars, $(Z_0, Z, \mu, A_t, A_\psi)$, and three one-forms, $(\omega_t, \omega_\psi, a)$, defined on the three-dimensional base whose metric is denoted by $ds_3^2$. In addition, we choose the convention for the five-dimensional Hodge star such that $\star_5 1=Z_0Z d \psi \wedge d t \wedge \mathrm{vol}_3$, where $\mathrm{vol}_3$ denotes the volume form on the three-dimensional base.

The three one-forms $(\omega_t, \omega_\psi, a)$ can be dualized into three scalars $(\Omega_t,\Omega_\psi,a)$ by imposing the Bianchi identities on the field strengths. As a result, the full solution is encoded in eight scalars on the three-dimensional base, according to
\begin{eqnarray}
\begin{split}
     d a+\mathcal{A}^{T} d \omega  =\tau^{-1}  \star \mathcal{B}, \qquad \mathcal{B} \equiv d B+ d \mathcal{A}^{^T} \varepsilon \mathcal{A}, \\
    \tau h d \omega  =\star \mathcal{V}, \qquad \mathcal{V}=d \Omega-3d B+ (d \mathcal{A}^{T} \varepsilon \mathcal{A}) \mathcal{A}   ,
\end{split}
\label{eq:selfduals}
\end{eqnarray}
where $\star$ is now the Hodge star operator on the three-dimensional base, and where we introduced
\begin{eqnarray}
\begin{split}
& h=Z^{-2}\left(\begin{array}{cc}
-1 & -\mu \\
-\mu &\quad \frac{Z^3}{Z_0}-\mu^2
\end{array}\right), \quad \tau=-\operatorname{det} h=\frac{1}{Z_0Z}, \\
& \omega=\binom{\omega_t}{\omega_\psi}, \quad \Omega=\binom{\Omega_t}{\Omega_\psi}, \quad \mathcal{A}=\binom{A_t}{A_\psi+\mu A_t}, \quad \varepsilon=\left(\begin{array}{cc}
0 & -1 \\
1 & 0
\end{array}\right) .
\end{split}
\label{eq:definitions}
\end{eqnarray}
The reduced three-dimensional action can then be written in terms of an $8\times 8$ coset matrix $\mathcal{M}$ that nontrivially depends on the eight scalars (see \eqref{app:MatrixM}) as
\begin{equation}
\mathcal{S}_3=\frac{1}{16 \pi G_3} \int\left(R_3 \star 1+\frac{1}{8} \operatorname{Tr}\left[d \mathcal{M}^{-1} \wedge \star d \mathcal{M}\right]\right) .
\end{equation}

It is also useful to introduce the dual matrix-valued one-form
$$
d \mathcal{N} \equiv \mathcal{M}^{-1} \star d \mathcal{M},
$$
which contains the information associated with the one-forms $(\omega_t, \omega_\psi, a)$.

The scalar target space of the coset model has a $G_{2(2)}$ symmetry, which can be represented in terms of the $\operatorname{SO}(4,4)$ isometry group of STU supergravity \cite{Chakraborty:2025ger}. In this $8\times 8$ matrix representation, the group is generated by 14 generators: two $\mathcal{H}$ generators, $\mathcal{H}_\pm$, two $\mathcal{P}$ generators, $\mathcal{P}_\pm$, two $\mathcal{W}$ generators, $\mathcal{W}_\pm$, two $\mathcal{Z}$ generators, $\mathcal{Z}_\pm$, two $\mathcal{X}$ generators, $\mathcal{X}_\pm$, and four generators $(\mathcal{O}_{\pm1},\mathcal{O}_{\pm2})$.

Further details on the sigma model are provided in Appendix~\ref{app:5d}, including the explicit expression of the matrix $\mathcal{M}$ in terms of the eight scalars, the equations of motion, and the embedding of the $G_{2(2)}$ subgroup in terms of the 14 generators of $\mathfrak{so}(4,4)$ listed above.

Finally, the sigma-model action is invariant under $G_{2(2)}$ transformations,
\begin{equation}\label{G22}
  \mathcal{M} \rightarrow g^T \mathcal{M} g, \quad \mathcal{N} \rightarrow g^{-1} \mathcal{N} g, \quad g \in G_{2(2)}.  
\end{equation}
These transformations therefore, generate new five-dimensional supergravity solutions from a known seed through simple algebraic matrix manipulations. In particular, they have been used to add charges to static and rotating black holes \cite{Cvetic:1995kv,Chow:2014cca}, as well as to smooth horizonless geometries \cite{Chakraborty:2025ger,Heidmann:2025pbb}. They can also be used to modify the asymptotic structure of a solution, through so-called subtraction transformations, for instance, to map an asymptotically flat solution to an asymptotically AdS one \cite{Cvetic:2011hp,Sakamoto:2025jtn}. Both types of transformations will be used here, and we will review them in the next section.

\subsection{Substraction and charging transformations} \label{subdisc}

There is a large variety of transformations that can be used to generate new solutions from a given seed. However, only a restricted subset leads to physically meaningful operations. In this paper, we focus on two classes of transformations: the \emph{subtraction transformations}, which change the asymptotics of a solution from flat to AdS$\times$S (and vice versa), and the \emph{(dis)charging transformations}, which convert the gravitational monopole charge into an electromagnetic charge, thereby allowing one to charge or discharge a solution.

\begin{itemize}

\item \underline{Charging transformations:}
There are three commuting subgroups of $G_{2(2)}$ transformations that act on the charges of a solution while preserving its asymptotics: the $\cP$-, $\cZ$-, and $\cW$-groups \cite{Chakraborty:2025ger}:
    \begin{equation}\label{disc}
        \begin{split}
        g_\cP &= e^{\delta_1 (\cP_+ + \cP_-) +\delta_2 (\cX_+ +\cX_-)}, \qquad g_\cZ = e^{\gamma_1 (\cZ_+ + \cZ_-) +\gamma_2 (\cO_{+1} +\cO_{-1})}, \\ 
        g_\cW &= e^{\alpha_1 (\cW_+ - \cW_-) +\alpha_2 (\cO_{+2} -\cO_{-2})}.
        \end{split}
    \end{equation}
    The explicit expressions of each $8\times 8$ matrix generator can be found in  Appendix \ref{App:GenerExpl}. The parameters $\delta_1$ and $\gamma_1$ generate, respectively, electric and magnetic charges of the U(1) gauge field. The parameters $\delta_2$ and $\gamma_2$ are associated with a boost and a KKm charge along the $\psi$ direction, while $\alpha_1$ and $\alpha_2$ generate dyonic charges.

    \item \underline{Subtraction transformations:}
    Subtraction transformations consist of appropriately modifying the asymptotic values of the fields. In this paper, we focus on several transformations that allow one to change the asymptotics from flat to AdS and back.
    \begin{itemize}
        \item[$\star$] $\IR^{1,3}\times$S$^1\longleftrightarrow$ AdS$_2\times$S$^3$:
       The subtraction transformation that maps $\IR^{1,3}\times$S$^1$ to AdS$_2\times$S$^3$ is
        \begin{equation}\label{gsz}
            g_{s\cZ}(d_1,d_2)=e^{d_1 \cZ_- +d_2 \cO_{-1}},
        \end{equation}
        where $d_1$ and $d_2$ are constants that must be fixed according to the asymptotic values of the metric components and the gauge potentials. For typical configurations, one usually requires $d_1=-d_2=1$. The reverse transformation from AdS$_2\times$S$^3$ to $\IR^{1,3}\times$S$^1$ is simply $g_{s\cZ}(d_1,d_2)^{-1}=g_{s\cZ}(-d_1,-d_2)$. Note that these transformations require the seed solution to have a nontrivial monopole source in $g_{tt}$ to function.\footnote{For instance, the Euclidean Schwarzschild geometry obtained from double Wick rotation of Schwarzschild$\times$S$^1$ is invariant under such transformations.}
        \item[$\star$] $\IR^{1,3}\times$S$^1\longleftrightarrow$ AdS$_3\times$S$^2$:
        The subtraction transformation that maps $\IR^{1,3}\times$S$^1$ to AdS$_3\times$S$^2$ is
        \begin{equation}
            g_{s\cP}(d_1,d_2)=e^{d_1 \cP_- +d_2 \cX_{-}},
            \label{gsp}
        \end{equation}
        where $d_1$ and $d_2$ must be similarly fixed depending on the asymptotic values of fields. For typical configurations, one usually has $d_1=d_2=1$. The reverse transformation from AdS$_3\times$S$^2$ to $\IR^{1,3}\times$S$^1$ is $g_{s\cP}(d_1,d_2)^{-1}=g_{s\cP}(-d_1,-d_2)$. These transformations require the seed solution to have a nontrivial monopole source in $g_{tt}$ or $g_{\psi\psi}$ to function.
        \item[$\star$] $\IR^{1,4}\longleftrightarrow$ AdS$_2\times$S$^3$:
       The subtraction transformation that maps $\IR^{1,4}$ to AdS$_2\times$S$^3$ is identical to $\IR^{1,3}\times$S$^1$ case, but without the $\cO_{-1}$ generator, which acts on the KKm sector. The transformation therefore reduces to $g_{s\cZ}(d_1,0)$. As before, $d_1$ is fixed by the asymptotic values of the fields, and typically $d_1=1$. The inverse transformation back to $\IR^{1,4}$ is simply $g_{s\cZ}(-d_1,0)$.
    \end{itemize}
\end{itemize}

\section{Generating rotation in asymptotically $\mathbb{R}^{1,3}\times \mathrm{S}^1$ spacetimes}
\label{sec:KerrGen}

In this section, we introduce a novel technique for generating rotation in four-dimensional geometries using purely algebraic sigma-model transformations in five-dimensional minimal supergravity.

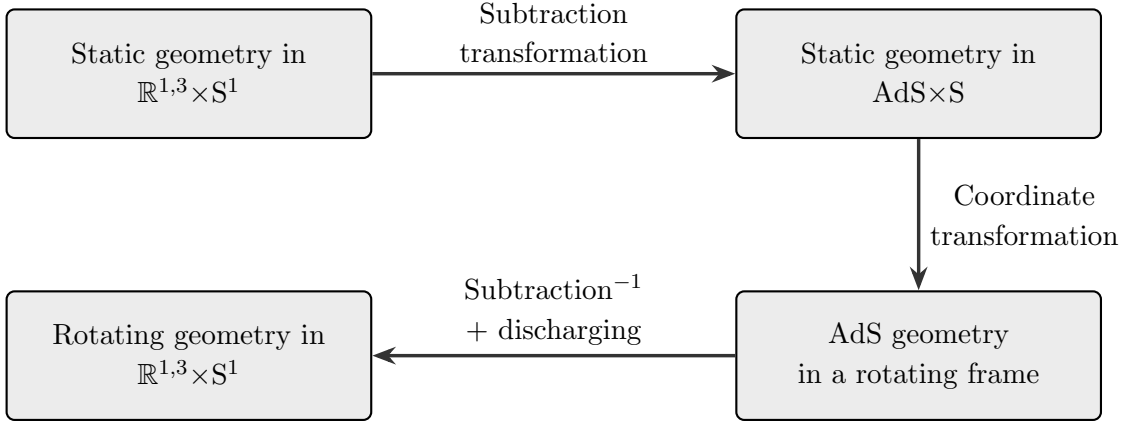
\begin{figure}[t]
\centering
\begin{tikzpicture}[
    >=Stealth,
    node distance=3.4cm and 4.8cm,
    box/.style={
        draw,
        rounded corners=3pt,
        thick,
        fill=gray!15,
        text width=4.4cm,
        minimum height=1.7cm,
        align=center,
        inner sep=6pt
    },
    arrow/.style={
        ->,
        line width=1.3pt,
        draw=black!80
    }
]

\node[box] (box1)
{Static geometry in\\ $\mathbb{R}^{1,3}\times$S$^1$};

\node[box, right=of box1] (box2)
{Static geometry in\\ AdS$\times$S};

\node[box, below=2cm of box2] (box3)
{AdS geometry\\ in a rotating frame};

\node[box, left=of box3] (box4)
{Rotating geometry in\\ $\mathbb{R}^{1,3}\times$S$^1$};

\draw[arrow]
(box1) -- node[above,align=center]
{Subtraction\\transformation} (box2);

\draw[arrow]
(box2) -- node[right,align=center]
{Coordinate\\transformation} (box3);

\draw[arrow]
(box3) -- node[above,align=center]
{Subtraction$^{-1}$ \\ + discharging} (box4);

\end{tikzpicture}

\caption{Sigma-model technique to systematically generate rotations for asymptotically $\mathbb{R}^{1,3}\times$S$^1$ spacetimes.}
\label{fig:genmap5d}
\end{figure}

The basic mechanism proceeds in three steps, as illustrated in Fig.~\ref{fig:genmap5d}. We begin with a static solution whose metric is asymptotically $\mathbb{R}^{1,3}\times$S$^1$. We then apply a \textit{subtraction transformation} that maps the asymptotic geometry either to AdS$_2\times$S$^3$ or to AdS$_3\times$S$^2$. Next, we move to a uniformly rotating frame by shifting the azimuthal coordinate by a term proportional to time. Finally, we apply the inverse \textit{subtraction transformation} to restore asymptotically flat boundary conditions. Under this inverse transformation, the rotation of the AdS frame translates into a genuine physical rotation in the asymptotically flat geometry. In general, the resulting background carries residual electric and magnetic charges; when necessary, these can be removed by applying appropriate \textit{discharging transformations}.

In the following subsections, we discuss these steps in detail and illustrate the procedure by applying it to the Schwarzschild black hole.

\subsection{The rotation-generating technique}

We start from a static geometry that is asymptotically $\mathbb{R}^{1,3}\times$S$^1$ and that admits the decomposition \eqref{eq:Ansatz}, where $\psi$ parametrizes the S$^1$ and $\phi$ is the azimuthal coordinate of the three-dimensional base. To generate rotation along the $\phi$ direction, we proceed as follows:

\begin{enumerate}

\item
\textit{Subtraction transformation:}
We first construct the coset matrices $\mathcal{M}$ and $\mathcal{N}$ associated with the seed geometry and act on them according to \eqref{G22} using either the group element $g=g_{s\mathcal{Z}}(1,-1)$ defined in \eqref{gsz} or $g=g_{s\mathcal{P}}(1,1)$ defined in \eqref{gsp}.\footnote{Depending on the asymptotic values of the seed fields, the transformation parameters may need to be adjusted.} The first transformation maps the asymptotic geometry to AdS$_2\times$S$^3$, while the second maps it to AdS$_3\times$S$^2$. Note that even when the seed solution carries no gauge field, the subtraction transformation generically activates a non-trivial background gauge field supporting electromagnetic charges.

The effect of this transformation is analogous to taking a decoupling limit of an asymptotically flat geometry.

\item
\textit{Coordinate transformation:}
We then shift the azimuthal coordinate to move to a uniformly rotating frame,
\begin{equation}
    \phi \to\phi+\Omega\, t,
    \label{eq:phiShift}
\end{equation}
where $\Omega$ is an arbitrary constant. After performing this transformation, we reconstruct the corresponding coset matrices $\mathcal{M}$ and $\mathcal{N}$ for the solution expressed in the rotating coordinate system.

\item
\textit{Inverse subtraction transformation:}
We next apply the inverse subtraction transformation using either the group element $g_{s\mathcal{Z}}(-1,1)$ or $g_{s\mathcal{P}}(-1,-1)$, depending on the choice made in the first step.\footnote{Once again, the subtraction parameters may differ from $\pm1$ and must sometimes be adjusted to match the asymptotic values of the fields.} This transformation restores the asymptotic structure to $\mathbb{R}^{1,3}\times$S$^1$, while preserving the rotation introduced by the coordinate shift. What was initially a frame-rotation artifact in the AdS region thus becomes a genuine physical rotation when the geometry is mapped back to asymptotically flat space. The resulting solution is therefore rotating along the S$^2$, with an angular momentum proportional to $\Omega$.

\item
\textit{Discharging transformation:}
The geometry obtained after the inverse subtraction transformation typically carries non-vanishing charges. When required, these charges can be removed by applying appropriate discharging transformations, $g_{\mathcal{Z}}(\gamma_1,\gamma_2)$ or $g_{\mathcal{P}}(\delta_1,\delta_2)$, defined in \eqref{disc}.

\end{enumerate}

It is important to emphasize that several continuous parameters enter this procedure and can affect the final geometry. For instance, constants added to the gauge potentials at intermediate steps can influence the action of the subtraction transformations. These parameters must therefore sometimes be chosen carefully to obtain the desired physical properties of the final solution.

Moreover, while the final discharging transformation guarantees to end up with a neutral solution, it does not guarantee to have a vacuum solution, as residual electromagnetic multipoles can a priori survive.

\subsection{An example: generating Kerr from algebraic transformations}

We begin by embedding the four-dimensional Schwarzschild metric into five dimensions:
\begin{equation}
ds^2=-fdt^2+\frac{dr^2}{f}+r^2 \left( d\theta^2+\sin^2 \theta \,d\phi^2 \right)+d\psi^2.
\label{eq:Schwarzschild4d}
\end{equation}
where $f=1-\frac{2m}{r}$. This metric can be cast into the sigma-model ansatz \eqref{eq:Ansatz}, with only two non-trivial scalar fields, $Z_0=Z=f^{-1/2}$. The corresponding coset matrices $\cM$ and $\mathcal{N}$ then follow directly from the general expressions \eqref{app:MatrixM} and \eqref{app:MatrixN}, and are given in \eqref{eq:SchwarzschildMandN}.

\begin{itemize}
    \item 
    {\it Subtraction transformation:} As discussed in the previous section, rotation can be generated through either the AdS$_3$ or AdS$_2$ route. For the Schwarzschild metric, both procedures ultimately lead to the same rotating solution. Here, we adopt the AdS$_2$ construction by applying the subtraction transformation $g=g_{s\cZ}(1,-1)$ \eqref{gsz} to the seed metric. The resulting geometry describes a non-extremal, static black hole in AdS$_2\times$S$^3$:
\begin{align}
    ds^2&=-\frac{ f r^2}{4 m^2}d t^2+\frac{4  m^2}{f r^2}d r^2+(d \psi-2 m \cos{\theta}\, d \phi )^2+4 m^2\left(d \theta^2+ \sin^2{\theta}\,d \phi^2 \right), \nn\\
    A=&\left(1-\frac{r}{2m}\right) \,dt. \label{eq:AdS2BH}
\end{align}
A noteworthy feature is that the electric potential possesses an asymptotic constant. This asymptotic value must be preserved until the inverse subtraction is performed at the very last step, ensuring that the final geometry asymptotes to flat $\IR^{1,3}\times$S$^1$. Similarly, the electric dual of the KK gauge field, $\Omega_\psi$ \eqref{eq:selfduals}, which enters the coset matrix $\mathcal{M}$, has an asymptotic value of $-1$ after this first transformation. This value must also be maintained throughout the intermediate steps.

\item
\textit{Coordinate transformation:} 
We now perform the shift $\phi\to\phi+\Omega\, t$, which brings the AdS$_2$ solution to a uniformly rotating frame:
\begin{align}
    ds^2= &-\frac{\Delta}{4m^2}\left(dt+\frac{8m^3\Omega \cos \theta}{\Delta}\left( d\psi-\frac{2mr^2 f \cos \theta}{\Sigma} d\phi\right) - \frac{16 m^4 \Omega \sin^2 \theta}{\Sigma} d\phi\right)^2\nn\\
    & +\frac{\Sigma}{\Delta} \left(d\psi-\frac{2mr^2 f\cos \theta}{\Sigma} d\phi \right)^2 +4m^2 \left[\frac{dr^2}{r^2 f}+d\theta^2+\frac{r^2 f\sin^2\theta}{\Sigma}d\phi^2 \right],
\end{align}
where we have defined $\Delta=r^2 f-16m^4 \Omega^2$ and $\Sigma=r^2 f-16m^4 \Omega^2 \sin^2 \theta$. This transformation generates the desired $dt\,d\phi$ cross-term, corresponding to rotation, which will be carried back to asymptotically flat space.

We then compute the coset matrices $\mathcal{M}$ and $\mathcal{N}$, which constitutes the most technically involved step. This requires determining the scalar duals of the one-forms $(B,\Omega_t,\Omega_\psi)$ using \eqref{eq:selfduals}, while ensuring that their asymptotic values remain $(0,0,-1)$, as fixed after the initial subtraction.

\item 
    \textit{Inverse subtraction transformation}: 
We next apply the inverse transformation $g=g_{s\cZ}(-1,1)$ to return to asymptotically flat space. This produces a Kerr-Newman black hole embedded in five-dimensional minimal supergravity, originally constructed in \cite{Cvetic:1996xz}, albeit in a different parametrization.
    
    \item 
    \textit{Discharging transformation}:
To recover the Kerr geometry, the electric and KKm charges must be removed. This is achieved via a discharging transformation $g_{\mathcal{Z}}(\gamma_1,\gamma_2)$. While such transformations typically eliminate only the monopole component of the gauge field, in the present case they remove the gauge potentials entirely.

Requiring the charges to vanish fixes the parameters to
\begin{equation}
    \gamma_1=-\gamma_2=\frac{1}{4} \log (1+16\Omega^2 m^2). 
\end{equation}

The resulting background corresponds to four-dimensional Kerr$\times$S$^1$, expressed in a non-standard parametrization. It can be brought to the Boyer-Lindquist form through the redefinitions
\begin{equation}
    m=\sqrt{M^2-a^2},\qquad \Omega=\frac{a}{4(M^2-a^2)},\qquad r\to r-M+\sqrt{M^2-a^2}\,.
\end{equation}

\end{itemize}

This example demonstrates that rotation can be introduced into any four-dimensional geometry in $\IR^{1,3}\times$S$^1$ through a sequence of purely algebraic transformations, without solving differential equations. The only step that may depend sensitively on the seed solution is the discharging transformation. In the present case, it yields a purely vacuum solution, whereas for more general seeds, one expects residual electromagnetic fields, typically appearing at large distances as dipole contributions.

\subsection{Rotating Weyl solutions}
\label{sec:RotWeyl}
To move beyond generating Kerr, we can test our rotation-generating technique on Weyl solutions \cite{Weyl:book,Israel1964}. By assuming axial symmetry and staticity, Weyl showed that the vacuum Einstein equations reduce to a linear Laplace equation on a flat two-dimensional space. As a result, they admit closed-form solutions in which the fields are sourced by rods along the symmetry axis. In four dimensions, these solutions necessarily describe a chain of Schwarzschild black holes separated by struts \cite{Israel1964,Costa:2000kf}, which can be interpreted as strings with negative tension. In higher dimensions, with additional compact directions, the same integrable structure persists \cite{Emparan:2001wk}; however, it allows for a much richer variety of solutions, including chains of bubbles and black holes \cite{Elvang:2002br,Bah:2020pdz}, as well as entirely smooth horizonless geometries \cite{Bah:2021owp,Bah:2021rki}. 

In this section, we restrict attention to the original four-dimensional Weyl solutions for simplicity and apply our technique to construct a rotating generalization of the initial ansatz. More precisely, we obtain configurations based on a linear Laplace equation with charged and rotating rod sources. This provides, to our knowledge, the first linear ansatz for nonextremal rotating and charged solutions of Einstein's equations.

In four dimensions, Weyl solutions take the form
\begin{equation}
    ds_4^2 = - U(\rho,z) \,dt^2 + \frac{1}{U(\rho,z)} \left[e^{2\nu(\rho,z)} (d\rho^2 +dz^2) +\rho^2 \,d\phi^2 \right], \label{eq:WeylMetric}
\end{equation}
where $\log U$ satisfies the Laplace equation on the flat $(\rho,z)$ space,
\begin{equation}
    (\partial_\rho^2+\partial_z^2)\log U=0\,, \label{eq:WeylEq}
\end{equation}
and $\nu$ is determined by first-order equations \cite{Israel1964,Emparan:2001wk}. To implement our method, we embed this ansatz trivially in five dimensions by adding a flat $\psi$ direction.

Since our construction involves dual one-forms, we introduce two additional fields $(V,W)$ derived from $U$ via the relations
\begin{equation}
    d(V\,d\phi) = \star_3 d \log U,\qquad dW= z^2 d\left(\frac{V}{z} \right)-\rho^2 d\log U\,, \label{eq:WeylDual}
\end{equation}
where $\star_3$ denotes the Hodge dual on the three-dimensional base space.

Applying all steps of the procedure, except for the discharging transformation, which is not essential for generating rotation, we obtain a five-dimensional solution that remains effectively four-dimensional, with $g_{\psi\psi}=1$. Upon reduction, this yields a solution of Einstein-Maxwell theory with a U(1) gauge field:\footnote{The five-dimensional embedding is $ds^2=(d\psi-\mathcal{A})^2+ds_4^2$, and the five-dimensional gauge field, $A$, is dual to the Kaluza-Klein gauge field, $\mathcal{A}$, $\star_4 dA= d\mathcal{A}$; the five-dimensional solution is given in Appendix \ref{app:RotWeyl}.}
\begin{equation}
\begin{split}
        ds_4^2 &= - \frac{U \Delta}{X^2+4\Omega^2 U S^2} (dt+\omega_t)^2 + \frac{X^2+4\Omega^2 U S^2}{U \Delta} \left[e^{2\nu} \Delta (d\rho^2 +dz^2) +\rho^2 \,d\phi^2 \right],\\
        \mathcal{A} &= \frac{\Omega\,\sqrt{U}\,(2S - \bar{U} V X)}{X^2 + 4 \Omega^2 U S^2} (dt+\omega_t)+\frac{\Omega^2\left( V^3 + \rho^2\bar{U}^2U^{-1}(2 S \sqrt{U} + V) \right)}{\Delta} d\phi, \label{eq:RotWeyl}
\end{split}
\end{equation}
with
\begin{align}
    \omega_t &\equi - \Omega \left(
2W + \frac{
2 V (V + z)
- \rho^2\bar{U}\,U^{-3/2}(1+U)
+ \Omega^2\left( T^2 - 2\rho^2\,\bar{U}^2 S (V \bar{U} - 2S) \right)
}{\Delta}
\right)\,d\phi, \nn\\
    \bar{U}&\equi \frac{U-1}{\sqrt{U}}\,, \qquad \Delta \equi 1-\rho^2 \Omega^2 \bar{U}^4\,,\qquad T \equi \rho^2 \bar{U}^2 +V^2, \\
    S&\equi \sqrt{U}V+z \bar{U},\qquad  X \equi 1-\Omega^2 T (U-1). \nn
\end{align}
Despite the presence of nontrivial gauge field and rotation one-forms, which typically introduce nonlinearity, the Einstein-Maxwell equations still reduce to the same linear Laplace equation for $\log U$ as in the static vacuum case \eqref{eq:WeylEq}. This establishes a linear ansatz for rotating and charged configurations in four dimensions.

\medskip

Regular Weyl solutions are specified by $N$ rods along the $z$-axis, each of length $M_i$, centered at $z=b_i$. Introducing adapted coordinates $(r_i,\theta_i)$ defined via the distances to the rod endpoints,
\begin{equation}
r_{ \pm}^{(i)} \equiv \sqrt{\rho^2+\left(z-(b_i\pm M_i)\right)^2},\quad r_i \equi \frac{r_+^{(i)}+r_-^{(i)}+2M_i}{2},\quad \cos \theta_i \equi \frac{r_-^{(i)}-r_+^{(i)}}{2M_i},
\label{eq:DistanceRods}
\end{equation}
the general solution reads\footnote{The most general solution is actually $U=\prod \left(1-\frac{2M_i}{r_i} \right)^{P_i}$, implying that the arguments in $V$ and $W$ are multiplied by $P_i$ while the arguments in $e^{2\nu}$ have a power $P_i P_j$. However, for $P_i\neq 1$, the rods are usually singular in four dimensions.}
\begin{align}
U &\= \prod_{i=1}^N \left(1-\frac{2M_i}{r_i} \right), \quad V \= \sum_{i=1}^N 2 M_i \,\cos \theta_i,\quad W \= - \sum_{i=1}^N 2 M_i \,(r_i-b_i \cos \theta_i)+c,\nn\\
e^{2\nu}&\=  \prod_{i,j=1}^N \frac{\left(\left(r_i-2M_i\right) \cos ^2 \frac{\theta_i}{2}+\left(r_j-2M_j\right) \sin ^2 \frac{\theta_j}{2}\right)\left(r_i \cos ^2\frac{\theta_i}{2}+r_j \sin ^2 \frac{\theta_j}{2}\right)}{\left(\left(r_i-2M_i\right) \cos ^2 \frac{\theta_i}{2} +r_j \sin ^2 \frac{\theta_j}{2}\right)\left(r_i \cos ^2 \frac{\theta_i}{2}+\left(r_j-2M_j\right) \sin ^2 \frac{\theta_j}{2}\right)}\,.\nonumber
\end{align}
In the present form, $\omega_t$ does not vanish asymptotically. Eliminating the NUT charge requires fixing the origin of the $z$-axis such that $\sum_{i=1}^N b_i M_i=0$, while removing the constant term fixes the constant in $W$, $c=4\left(\sum M_i^2-2(\sum M_i)^2(1+2\Omega^2 (\sum M_i)^2)\right)$.

We have verified that, after the transformation, the rods correspond to Kerr-Newman black holes aligned along the axis. However, as in other configurations of rotating black holes \cite{dietz1982stationary,101063,Breton_1995}, the segments between the black holes do not automatically correspond to degeneracies of the $\phi$-circle (i.e., struts). This would require $\omega_t=0$ along those segments, which cannot be achieved by tuning the parameters within our ansatz. As a consequence, these loci correspond to Misner strings, yielding a continuous chain of rotating and charged black holes prevented from collapse by Misner strings. 

Despite these pathological Misner strings, the construction demonstrates the effectiveness of our rotating-generating technique. To our knowledge, it provides the first linear ansatz for rotating and charged masses beyond the BPS ansatz, which is restricted to extremal configurations. We will propose direction to cure the presence of CTC in the next section.

\subsection{Potential new applications}
\label{sec:Application}

 The rotation-generating technique admits a broad range of applications across different classes of gravitational backgrounds that asymptote to $\mathbb{R}^{1,3}\times$S$^1$. In what follows, we outline several representative directions in which these methods can be applied, highlighting their potential utility and the scope of our technique.
\begin{itemize}
\item[$\blacklozenge$] The most natural generalization of our previous construction is to apply our rotation-generating technique to the Euclidean version of the vacuum seed considered above:
\begin{equation}
    ds^2 = - dt^2 +U \,d\psi^2+ \frac{1}{U} \left[e^{2\nu} (d\rho^2 +dz^2) +\rho^2 \,d\phi^2 \right],
\end{equation}
which includes, in particular, the Euclidean Schwarzschild geometry when $U$ is sourced by a single rod.

Since $g_{tt}$ is trivial, the construction requires following the AdS$_3$ route of our procedure. We have already analyzed the case where the seed is the Euclidean Schwarzschild geometry and find that the resulting solution corresponds to a subclass of the rotating topological stars constructed in \cite{Bianchi:2025uis}. However, this configuration exhibits tensions between regularity at the rod and the periodicity lattice required by the $\mathbb{R}^{1,3}\times$S$^1$ asymptotics.

The case where $U$ is sourced by multiple rods is therefore particularly interesting to investigate, as it may lead to (most probably singular) configurations describing interacting rotating topological stars.

    \item[$\blacklozenge$] It would also be interesting to explore the outcome of the procedure starting from the generalized Weyl geometries of \cite{Emparan:2001wk}:
    \begin{equation}
    ds^2 = - U_1\, dt^2 +U_2 \,d\psi^2+ \frac{1}{U_1 U_2} \left[e^{2\nu} (d\rho^2 +dz^2) +\rho^2 \,d\phi^2 \right], \label{eq:GenWeyl}
    \end{equation}
    where $\log U_1$ and $\log U_2$ satisfy linear Laplace equations.
    
This ansatz contains a wide variety of solutions, ranging from chains of black holes on Kaluza--Klein bubbles \cite{Costa:2000kf,Bah:2020pdz} to the five-dimensional embedding of the Zipoy--Voorhees metric with a scalar field \cite{Zipoy:1966btu,Voorhees:1970ywo,Stephani:2003tm,Kodama:2003ch}:
\begin{equation}
    ds^2 = -f^{\alpha}dt^{2}+\frac{1}{f^{-1+\alpha+\beta}}\left[\left(\frac{H}{f}\right)^{1-\alpha^{2}-\alpha\beta-\beta^{2}}\left(\frac{dr^{2}}{f}+r^{2}d\theta^{2}\right)+r^{2}\sin^{2}\theta d\phi^{2}\right]+f^{\beta}d\psi^{2}
\end{equation}
with $f=1-\frac{2m}{r}$ and $H=1-\frac{2m}{r}+\frac{m^2 \sin ^2\theta}{r^2}$.

Interestingly, the rotating geometries arising from \eqref{eq:GenWeyl} may resolve the pathology encountered in the rotating Weyl solutions \eqref{eq:RotWeyl}. In particular, the segments separating the black-hole rods could become regular coordinate degeneracies where a spacelike Killing vector of the form $\partial_\psi + \alpha \partial_\phi$ smoothly degenerates.

\item[$\blacklozenge$] A large variety of nonextremal static geometries beyond the traditional black hole solutions has been constructed in recent years. Examples include neutral bound states of branes and antibranes supported by smooth bubbles, reproducing half of the Schwarzschild entropy \cite{Heidmann:2023kry,Heidmann:2025yzd}, as well as entirely smooth horizonless geometries built from single or multiple nonextremal bubbles \cite{Bah:2020pdz,Bah:2021rki,Heidmann:2021cms,Bah:2022yji,Bah:2023ows}. This broad class of solutions provides a rich set of static seeds to which our rotation-generating algorithm can be applied in order to construct their rotating counterparts. In particular, generating rotating nonextremal smooth horizonless geometries has long been a major challenge. Existing solutions are either outside the black hole regime, such as the JMaRT geometry \cite{Jejjala:2005yu}, the rotating topological star \cite{Heidmann:2025pbb}, or the running bolt \cite{Bena:2009qv,Bena:2025usu}, or suffer from tensions between smoothness at the bolt and Kaluza--Klein asymptotics \cite{Bianchi:2025uis}. It will therefore be particularly interesting to apply our technique systematically to generate large families of rotating smooth geometries and investigate their physical properties.

\item[$\blacklozenge$] Finally, there also exist large families of nontrivial BPS and almost-BPS asymptotically $\IR^{1,3}\times$S$^1$ solutions depending on up to three coordinates, including multicenter black holes \cite{Denef:2000nb,Bates:2003vx,Goldstein:2008fq} and multicenter bubbling geometries \cite{Bena:2007kg,Heidmann:2017cxt,Bena:2017fvm}. Applying our rotation-generating technique to these solutions could lead to new classes of nonextremal geometries and provide a novel mechanism for breaking supersymmetry in a controlled and highly nontrivial way.

\end{itemize}


\section{Generating rotation in asymptotically $\mathbb{R}^{1,4}$ spacetimes}
\label{sec:MPGen}

Since we work in five dimensions, we can generalize the previous technique to geometries asymptotic to $\mathbb{R}^{1,4}$. While $\mathbb{R}^{1,3}\times$S$^1$ admits only a single independent rotation, solutions in $\mathbb{R}^{1,4}$ can generically carry two independent angular momenta, which requires a significant adaptation of the method. Nevertheless, the underlying principles remain the same: we use sigma-model transformations to move to AdS$_2\times$S$^3$ asymptotics, perform a time shift along one of the S$^3$ angular directions, and then return to asymptotically flat space.

In this section, we present this rotation-generating technique in detail, before applying it to the Schwarzschild-Tangherlini black hole, which yields the Myers-Perry black hole in a few algebraic steps.

\subsection{The rotation-generating technique}
\label{sec:GenTech5D}

For geometries asymptotic to $\mathbb{R}^{1,3}\times$S$^1$, the sigma-model decomposition is natural: the extra compact dimension is used for the dimensional reduction to three dimensions (to construct $(\mathcal{M},\mathcal{N})$ and implement the sigma-model transformations), while the azimuthal angle of the S$^2$ is in the three-dimensional base and used to generate rotation.

In contrast, for $\mathbb{R}^{1,4}$ geometries, the situation is less straightforward since the two compact directions are on equal footing. A three-sphere is typically described by the line element
\begin{equation}
    d\Omega_3^2 \= d\tau^2+\cos^2 \tau \,d\varphi_1^2+\sin^2 \tau \,d\varphi_2^2,\qquad 0\leq\tau\leq\frac{\pi}{2}. \label{eq:S3tradi}
\end{equation}
One might then attempt to select one angular coordinate, say $\varphi_2$, for the sigma-model reduction and use the other for the shift generating rotation. However, we observed this procedure does not produce a rotation purely along $\varphi_1$, but rather along a linear combination such as $\varphi_1 + \varphi_2$.

This suggests that it is more convenient to work in a basis adapted to the Hopf fibration of S$^3$, in which the three-sphere is described as an S$^1$ fiber over S$^2$:
\begin{equation}
    d\Omega_3^2 = \frac{1}{4}\left[ d\theta^2 +\sin^2 \theta \,d\phi^2 + \left(d\psi-\cos \theta d\phi \right)^2\right] = \frac{1}{4}\left[ d\theta^2 +\sin^2 \theta \,d\psi^2 + \left(d\phi-\cos \theta d\psi \right)^2\right],
    \label{eq:HopfDec}
\end{equation}
where $\theta = 2\tau$, $\phi=\varphi_1+\varphi_2$ and $\psi=\varphi_2-\varphi_1$.

In this parametrization, the analogy with the $\mathbb{R}^{1,3}\times$S$^1$ case becomes manifest. One can treat the $\psi$ coordinate (respectively $\phi$) as the reduction direction and use $\phi$ (respectively $\psi$) as the shift direction to generate rotation along $\phi$ (respectively $\psi$). By repeating the procedure for both choices, one can independently generate rotations along $\phi$ and $\psi$, thereby constructing the solution with two angular momenta in $\mathbb{R}^{1,4}$.

\begin{figure}[t]
\label{mpbh}
\centering
\begin{tikzpicture}[
    >=Stealth,
    box/.style={
        draw,
        rounded corners=3pt,
        thick,
        fill=gray!15,
        text width=4.5cm,
        minimum height=1.7cm,
        align=center,
        inner sep=6pt
    },
    arrow/.style={
        ->,
        line width=1.3pt,
        draw=black!80
    }
]


\node[box] (box1) at (0,0)
{Static geometry\\ in $\mathbb{R}^{1,4}$};

\node[box] (box2) at (9,0)
{Static geometry\\ in AdS$_2\times$S$_3$};

\node[box] (box3) at (9,-3.5)
{AdS$_2\times$S$_3$ geometry\\ in a rotating frame};

\node[box] (box5) at (9,-7)
{Rotating geometry\\ in AdS$_2\times$S$_3$};

\node[box] (box6) at (9,-10.5)
{Rotating AdS$_2\times$S$_3$ geometry\\ in a rotating frame};

\node[box] (box4) at (0,-5.2)
{Geometry in $\mathbb{R}^{1,4}$\\ with one rotation};

\node[box] (box7) at (0,-10.5)
{Geometry in $\mathbb{R}^{1,4}$\\ with two rotations};


\draw[arrow]
(box1.east) -- node[above, align=center]
{Subtraction\\transformation} (box2.west);

\draw[arrow]
(box2.south) -- node[right, align=center]
{Coordinate\\transformation} (box3.north);

\draw[arrow]
(box3.west)
 -- node[pos=0.5, above, align=center]{Subtraction$^{-1}$\\+ discharging}
 ++(-3.5,0)
 |- ($(box4.east)+(0,0.2cm)$);

\draw[arrow]
($(box4.east)+(0,-0.2cm)$) -| ++(0.55,-1.6)
-- node[pos=0.5, above, align=center]
{Refibration\\+ subtraction} (box5.west);

\draw[arrow]
(box5.south) -- node[right, align=center]
{Coordinate\\transformation} (box6.north);

\draw[arrow]
(box6.west) -- node[above, align=center]
{Subtraction$^{-1}$\\+ discharging} (box7.east);

\end{tikzpicture}

\caption{Schematic sequence of transformations used to generate rotations in asymptotically $\mathbb{R}^{1,4}$ spacetimes.
}
\label{fig:genmap5d2}
\end{figure}
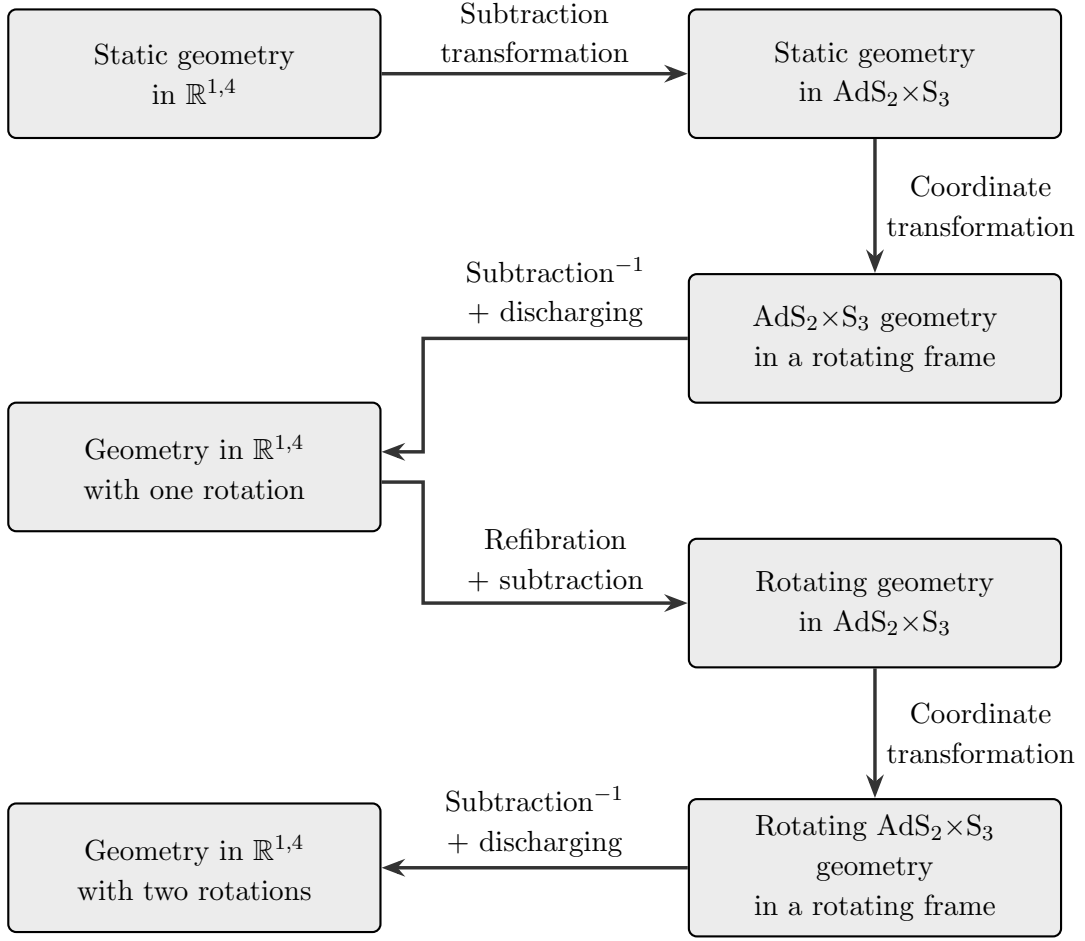

The sequence of steps is summarized in Fig.\ref{fig:genmap5d2}, which we now describe in detail. We start from a static geometry asymptotic to $\mathbb{R}^{1,4}$, where $\phi$ and $\psi$ denote the two Hopf angles of the S$^3$ as given in \eqref{eq:HopfDec}. To generate the two independent rotations on the S$^3$, we proceed as follows:

\begin{enumerate}
\item[(A)] \underline{Generating rotation along $\phi$:}
\end{enumerate}
\begin{enumerate}
\item 
\textit{Subtraction transformation:}

We begin by choosing $\psi$ as the fiber coordinate and $\phi$ as the base angle for the sigma-model decomposition described in \eqref{sec:SigmaModel}, and construct the corresponding coset matrices $\mathcal{M}$ and $\mathcal{N}$ for the seed geometry. We then act with the subtraction transformation $g_{s\mathcal{Z}}(1,0)$ defined in \eqref{gsz}, which maps the asymptotically flat geometry to an asymptotically AdS$_2\times$S$^3$ configuration.

\item
\textit{Coordinate transformation:}
We shift the azimuthal base coordinate $\phi$ to move to a uniformly rotating frame,
\begin{eqnarray}\label{phishift}
\phi \to \phi + \Omega_1 t\,,
\end{eqnarray}
where $\Omega_1$ is a constant. After performing this transformation, we reconstruct the corresponding coset matrices $\mathcal{M}$ and $\mathcal{N}$, still treating $\phi$ as the base angle and $\psi$ as the fiber.

\item
\textit{Inverse subtraction transformation and discharging transformation:}
We next apply the inverse subtraction transformation $g_{s\mathcal{Z}}(-1,0)$, which restores the asymptotically flat geometry $\mathbb{R}^{1,4}$ while preserving the rotation introduced by \eqref{phishift}. The resulting solution therefore carries angular momentum along $\phi$, proportional to $\Omega_1$.

As in the $\mathbb{R}^{1,3}\times S^1$ case, the solution generically also acquires electric charges under the U(1) gauge field. These can be removed by acting with the discharging transformation $g_{\mathcal{Z}}(\gamma,0)$ defined in \eqref{disc}, yielding a neutral asymptotically $\mathbb{R}^{1,4}$ geometry with a single rotation.
\end{enumerate}

\begin{enumerate}
\item[(B)] \underline{Generating rotation along $\psi$:}
\end{enumerate}
\begin{enumerate}
\item[4.]
\textit{Refibration and Subtraction transformation:}

We now rewrite the geometry by completing the fibration along the $\phi$ direction, treating $\psi$ as the base angle and constructing the corresponding coset matrices $\mathcal{M}$ and $\mathcal{N}$ for this new decomposition. We then apply the subtraction transformation $g_{s\mathcal{Z}}(1,0)$, bringing the geometry back to AdS$_2\times$S$^3$, but with $\phi$ and $\psi$ interchanged in their roles.

\item[5.]
\textit{Coordinate transformation:} We shift the azimuthal angle $\psi$,
\begin{eqnarray}
    \psi\to \psi +\Omega_2 t
\end{eqnarray}
and compute the corresponding matrices $\mathcal{M}$ and $\mathcal{N}$ in this uniformly rotating frame.

\item[6.]
\textit{Inverse subtraction transformation and discharging transformation:}
We apply the inverse subtraction transformation $g_{s\mathcal{Z}}(-1,0)$, restoring asymptotically flat $\mathbb{R}^{1,4}$ geometry. The resulting solution now carries two independent angular momenta associated with the two azimuthal directions of the S$^3$, proportional to $\Omega_1$ and $\Omega_2$, respectively.

As before, the geometry may also carry electromagnetic charges, which can be eliminated by applying the discharging transformation $g_{\mathcal{Z}}(\gamma,0)$.
\end{enumerate}

As for the $\IR^{1,3}\times$S$^1$ technique, it is important to emphasize that several continuous parameters enter this procedure and can affect the final geometry. For instance, constants added to the gauge potentials at intermediate steps can influence the action of the subtraction transformations. These parameters must therefore sometimes be chosen carefully to obtain the desired physical properties of the final solution.

Moreover, while the final discharging transformation guarantees to end up with a neutral solution, it does not guarantee to have a vacuum solution, as residual electromagnetic multipoles may survive. 

\subsection{An example: generating Myers-Perry from algebraic transformations}

In this section, we construct the Myers-Perry black hole \cite{Myers:1986un,Myers:2011yc} starting from the Schwarzschild-Tangherlini solution using the rotation-generating technique described above. We begin by reviewing the target geometry. 

In \cite{Myers:1986un,Myers:2011yc}, the rotating black hole is typically expressed in the $(\tau,\varphi_1,\varphi_2)$ parametrization of the S$^3$ \eqref{eq:S3tradi}, with two angular momentum parameters $(a,b)$ associated with rotations along $\varphi_1$ and $\varphi_2$, respectively. Since our construction is naturally adapted to Hopf-fibered coordinates $(\phi,\psi)$, we rewrite the solution in this coordinate system. We also introduce the combinations $\rma \equiv (a+b)/2$ and $\rmb \equiv (a-b)/2$, corresponding to rotations along $\phi$ and $\psi$, in terms of which the metric takes the form
\begin{equation}
\begin{split}
    ds^2 = & -\left(1-\frac{M}{\rho^2} \right) \left(dt+\frac{M(-\rmb+\rma \cos \theta)}{2(\rho^2-M)} d\psi+\frac{M(-\rma+\rmb \cos \theta)}{2(\rho^2-M)} d\phi\right)^2 \\
    & +\frac{\Sigma}{4(\rho^2-M)} \left(d\psi+\frac{\rma \rmb (2(r^2+\rma^2+\rmb^2)-M)\sin^2\theta-\Delta \cos \theta}{\Sigma}\,d\phi \right)^2  \\
    & + \frac{r^2 \rho^2}{\Delta} dr^2+ \frac{\rho^2}{4} \left(d\theta^2 + \frac{\Delta}{\Sigma} \sin^2 \theta \,d\phi^2 \right) 
\end{split}
\label{eq:MPGen}
\end{equation}
where we have defined
\begin{equation}
\begin{split}
    \Sigma&\equiv \Delta-\rma^2(M-4\rmb^2) \sin^2 \theta,\qquad \Delta\equiv (r^2+\rma^2+\rmb^2)^2-M r^2-4\rma^2\rmb^2,\\
    \rho^2 &\equiv r^2+\rma^2+\rmb^2+2 \rma \rmb \cos \theta.
\end{split}
\label{eq:Delt&Rho}
\end{equation}

In what follows, we refer to the Myers-Perry solution with two independent angular momenta as “MP2,” while the singly rotating case, obtained by setting $\rmb=0$ (i.e. rotation along $\phi$ only), will be denoted “MP1.”

\subsubsection{From Schwarzschild-Tangherlini to MP1}

We start with the five-dimensional Schwarzschild-Tangherlini solution \cite{Tangherlini:1963bw}: 
\begin{equation}
    ds^{2}
    =
    -fdt^{2}+\frac{dr^2}{f}
    +
    \frac{r^{2}}{4}\left[ d\theta^2 +\sin^2 \theta \,d\phi^2 + \left(d\psi-\cos \theta d\phi \right)^2\right] \, .
    \label{eq:SchwarzschildTan}
\end{equation}
where $f=1-\frac{m}{r^{2}}$. This metric can be cast into the sigma-model ansatz 
\eqref{eq:Ansatz}, 
with only two non-trivial scalar fields, $Z_0=4r^{-2}f^{-1/2},Z=f^{-1/2}$ and $\omega_\psi=-\cos\theta d\phi$. The corresponding coset matrices $\cM$ and $\mathcal{N}$ are given in \eqref{eq:SchwarzschildTMandN}.

\begin{itemize}
    \item 
    \textit{Subtraction transfromation:}
    We start by applying the subtraction transformation $g_{s\mathcal{Z}}(1,0)$  \eqref{gsz} to the seed metric. The resulting geometry is given by
    \begin{equation}
    \begin{split}
        ds^{2}
        &=
        -\frac{fr^{4}}{m^{2}}dt^{2}+\frac{m}{fr^{2}}dr^{2}+\frac{1}{4}m(d\psi-\cos\theta d\phi)^{2}+\frac{1}{4}m(d\theta^{2}+\sin^{2}\theta d\phi^{2}) \, , \\
        A
        &=
        \left(1-\frac{r^2}{m}\right) dt \, ,
        \end{split}
    \end{equation}
which is the same non-extremal static black hole in $\mathrm{AdS}_2 \times$S$^3$ as in \eqref{eq:AdS2BH} with a slightly different parametrization. The asymptotic value of the gauge potential is non-vanishing and 
must be preserved until the inverse subtraction transformation is applied, thereby ensuring that the resulting geometry asymptotes to flat $\mathbb{R}^{1,4}$.
    \item 
    \textit{Coordinate transformation:}
    We now shift $\phi\to\phi+\Omega_1t$, bringing the AdS$_2$ solution into a rotating frame:
    \begin{equation}
\begin{aligned}
d s^2= & -\frac{\Delta}{4 m^2}\left(d t+\frac{\Omega_1 m^3 \cos \theta}{\Delta}\left(d \psi-\frac{4 r^4 f \cos \theta}{\Sigma} d \phi\right)+\frac{\Omega_1 m^3 \sin ^2 \theta}{\Sigma} d \phi\right)^2 \\
& +\frac{\Sigma}{4}\left(d \psi-\frac{4 r^4 f \cos \theta}{\Sigma} d \phi\right)^2+\frac{m}{4}\left(\frac{4 d r^2}{r^2 f}+d \theta^2+\frac{4 r^4 f \sin ^2 \theta}{\Sigma} d \phi^2\right) \, , 
\end{aligned}
\end{equation}

    where we have defined $\Delta = 4 r^4-4 m r^2-\Omega_1^2 m^3$  and  $\Sigma = \Delta+\Omega_1^2 m^3 \cos ^2 \theta$. This transformation generates the $dt\,d\phi$ cross-term, responsible for generating the rotation when transformed back to asymptotically flat space.
    \item
    \textit{Inverse subtraction $\&$ discharging:}
    We now apply the inverse subtraction transformation $g_{s\mathcal{Z}}(-1,0)$ back to asymptotically flat spacetime. Thus produces a charged rotating black hole solution in five-dimensional minimal supergravity with a single rotation, also called the Cvetic-Youm black hole \cite{Cvetic:1996xz}.
  
    To obtain Myers-Perry solution with one angular momentum, the electric charges must be removed. This is achieved using the discharging transformation $g_{\mathcal{Z}}(\gamma,0)$. While such transformations typically eliminate only the monopole component of the gauge field, in this case, they remove the gauge potentials entirely.

Requiring the charges to vanish fixes the parameters to
\begin{equation}
    \gamma
    =
    \frac{1}{4} \log (1+ \Omega_1^2 m)~.
\end{equation}

As a result, we obtain MP1 in slightly different coordinates and parametrization as in \eqref{eq:MPGen}. To cast the metric in the standard form, one performs the following identifications:
\begin{align}
    m 
    = 
    \sqrt{M(M-\rma^2)} 
    \, , 
    \quad 
    \Omega_1 
    = 
    \frac{\rma}{M^{\frac{1}{4}}(M-\rma^2)^{\frac{3}{4}}}
    \, , \quad 
    r^2 
    \to 
    r^2+\frac{\rma^2}{4} 
    -
    \frac{M-\sqrt{M(M-\rma^2)}}{2} . \nn
\end{align}


\end{itemize}

\subsubsection{From MP1 to MP2}
Above, we obtained the $\mathrm{b=0}$ Myers-Perry black hole \eqref{eq:MPGen} starting from the Schwarzschild- Tangherlini solution. We now generate the remaining rotation by using this geometry as a seed.

\begin{itemize}
    \item 
    \textit{Refibration $\&$ Subtraction transformation}: As discussed in Section \ref{sec:GenTech5D}, generating rotation along $\psi$ requires viewing MP1 as a $(t,\phi)$ fibration over a $(r,\theta,\psi)$ base, in order to construct the associated sigma model. In this decomposition, the MP1 geometry takes the form
    \begin{align}
        ds^{2}=&-\left(1-\frac{M}{\rho^{2}}\right)\left(dt-\frac{\mathrm{a}M}{2(\rho^{2}-M)}(d\phi-\cos\theta d\psi)\right)^{2}\\&+\frac{\Delta}{4(\rho^{2}-M)}(d\phi-\cos\theta d\psi)^{2}+\frac{r^{2}\rho^{2}}{\Delta}dr^{2}+\frac{\rho^{2}}{4}(d\theta^{2}+\sin^{2}\theta d\psi^{2})\,, \nn
    \end{align}
    where $\rho$ and $\Delta$ are defined in \eqref{eq:Delt&Rho} with $\rmb=0$. 

    We then construct the coset matrices $\mathcal{M}$ and $\mathcal{N}$ associated with this decomposition using \eqref{app:MatrixM} and \eqref{app:MatrixN}, and act on them with the group element $g = g_{s\mathcal{Z}}(1,0)$. This produces a rotating nonextremal black hole with AdS$_2 \times$S$^3$ asymptotics:
    \begin{equation}
    \begin{aligned}
    ds^{2}
    &=-
    \frac{\rho^{2}(\rho^{2}-M)}{M^{2}}\left[dt-\frac{\mathrm{a}M}{2}\frac{2\rho^{2}-M}{\rho^2(\rho^{2}-M)}(d\phi-\cos\theta d\psi)\right]^{2}
    \\
    &+
    \frac{M\Delta}{4\rho^{2}(\rho^{2}-M)}(d\phi-\cos\theta d\psi)^{2}+\frac{Mr^{2}}{\Delta}dr^{2}+\frac{M}{4}\left(d\theta^{2}+\sin^{2}\theta d\psi^{2}\right) \, , 
    \\
    A&=\left(1-\frac{\rho^2}{M}\right) dt + \mathrm{a} (d\phi-\cos{\theta}d\psi ) \, , 
    \end{aligned}
    \end{equation}
    where
    \begin{equation}
        \rho^2=r^2+\mathrm{a}^2\,,\quad \Delta =(r^2+\mathrm{a}^2)^2-Mr^2~.
    \end{equation}
    Several constructions of non-extremal rotating black holes in AdS$_2\times$S$^3$, AdS$_2\times$S$^2$, and AdS$_3\times$S$^2$ exist in the literature \cite{Cvetic:2011hp,Virmani:2012kw,Sahay:2013xda}. All of them are obtained via subtraction transformations applied to asymptotically flat black holes. However, to our knowledge, the above solution is a new rotating geometry in AdS$_2\times$S$^3$, carrying both electric and magnetic charges.

    \item
    \textit{Coordinate transformation}:  
    We shift the coordinate $\psi$ as $\psi \to \psi + \Omega_2 t$, bringing the geometry to a uniformly rotating frame along $\psi$. We then derive the coset matrices $\mathcal{M}$ and $\mathcal{N}$ in this new basis.

    \item 
    \textit{Inverse subtraction transformation and discharging transformation}: We return to asymptotically flat space by applying the inverse subtraction transformation $g^{-1} = g_{s\mathcal{Z}}(-1,0)$, obtaining the Cvetič--Youm black hole with two angular momenta and equal electric charges \cite{Cvetic:1996xz}, in a slightly different parametrization.

To recover the Myers--Perry solution, the electric charges must be removed. This is achieved using the discharging transformation $g_{\mathcal{Z}}(\gamma,0)$. Requiring the charges to vanish fixes the parameter to
    \begin{equation}
        \gamma
        =
        \frac{1}{4}
        \log{(1+\Omega_2^2 M)} 
        \, ,
    \end{equation}
which also makes the whole gauge potential vanish.

The resulting geometry corresponds to MP2, but in coordinates and parameters different from those in \eqref{eq:MPGen}. To bring the metric to that form, one performs:
\begin{equation}
\begin{aligned}
    M 
    &= 
    \sqrt{m^2-4m\mathrm{b}^2}, 
    \quad 
    \Omega_2 
    = 
    -\frac{2\mathrm{b}\sqrt{m}}
    {[m^2-4m\mathrm{b}^2]^{\frac{3}{4}}} \, , \quad
    \mathrm{a} 
    \to 
    -\mathrm{a}\left[1-\frac{4\mathrm{b}^{2}}{m}\right]^{\frac{1}{4}} \, ,
    \\
    r^{2}
        &
        \to
        r^2+\frac{2(\mathrm{a}^2+\mathrm{b}^2)- m}{2}+\frac{ m-2\mathrm{a}^2}{2 m} \sqrt{m\left(m-4\mathrm{b}^2\right)}  ~.  \\ 
\end{aligned}
\end{equation}

\end{itemize}

This completes the construction of the Myers-Perry black hole from the Schwarzschild-Tangherlini solution using our rotation-generating technique, which relies only on algebraic operations and simple transformations. The only subtlety arises in the final step, where control over the gauge fields is limited. While in the Myers-Perry case neutralization yields a vacuum solution, there is no guarantee that applying the same procedure to other geometries eliminates all electromagnetic contributions. In general, one may obtain neutral geometries that still carry nontrivial electromagnetic fluxes, inducing dipole or higher multipole moments.

\subsection{Potential new applications}
\label{sec:Application2}

The rotation-generating technique developed in the previous subsection admits a broad range of applications across gravitational backgrounds asymptotic to $\mathbb{R}^{1,4}$. In the following, we outline several representative directions where this method may lead to interesting results.

\begin{itemize}
\item[$\blacklozenge$] First, the absence of a uniqueness theorem in five-dimensional asymptotically flat gravity allows for a much richer landscape of black objects than in four dimensions. Beyond the traditional spherical black hole, one may consider, for instance, black rings \cite{Emparan:2006mm,Pomeransky:2006bd} and black Saturn configurations \cite{Elvang:2007rd}. These geometries necessarily rely on rotation to balance the gravitational attraction and ensure regularity. Nevertheless, one can still consider their static and singular limits as seed geometries and investigate the configurations generated by our rotation-generating algorithm.

\item[$\blacklozenge$] Second, the static Weyl ansatz \eqref{eq:GenWeyl} also allows for geometries asymptotic to $\mathbb{R}^{1,4}$. This can be achieved by sourcing $U_2$ with a semi-infinite rod \cite{Emparan:2001wk}. Equivalently, the ansatz can be rewritten in a form where the asymptotic S$^3$ appears directly in Hopf-fibered coordinates, as discussed in \cite{Heidmann:2021cms}. As in the $\mathbb{R}^{1,3}\times$S$^1$ case, it would therefore be natural to apply our algorithm to these Weyl solutions and investigate whether it yields a linear family of geometries describing nonextremal rotating and charged sources in five dimensions.

\item[$\blacklozenge$] Finally, similarly to the $\mathbb{R}^{1,3}\times$S$^1$ geometries discussed previously, there exists a large class of BPS smooth horizonless solutions depending on at most three coordinates to which our construction may be applied. Besides the multicenter geometries already mentioned, one may consider the two-charge supertube solutions \cite{Lunin:2001fv} as well as the $(1,0,n)$ superstrata \cite{Bena:2015bea,Bena:2017xbt,Heidmann:2019xrd}, which admit consistent truncations to five-dimensional supergravity. These backgrounds provide particularly interesting laboratories to investigate how the rotation-generating procedure acts on supersymmetric microstate geometries and whether it can produce new families of rotating nonextremal horizonless solutions.

\end{itemize}

\section{Discussion}
\label{sec:Discussion}

In this paper, we have constructed a purely algebraic technique to generate rotation for arbitrary solutions asymptotic to $\mathbb{R}^{1,3}\times$S$^1$ or $\mathbb{R}^{1,4}$ in five-dimensional minimal supergravity. The procedure consists of mapping a static seed geometry to an AdS$\times$S background through sigma-model transformations, performing a time-dependent coordinate shift to move to a rotating frame, and subsequently returning to asymptotically flat spacetime by applying the inverse sigma-model transformation. To demonstrate the efficiency and versatility of this algorithm, we recovered the Kerr and Myers--Perry black holes from the four-dimensional Schwarzschild and five-dimensional Schwarzschild--Tangherlini solutions respectively, and constructed new rotating linear solutions in $\mathbb{R}^{1,3}$ by applying the method to static Weyl geometries. Our construction therefore provides a powerful new framework for systematically generating rotating solutions in supergravity.

Beyond the representative applications discussed in Sections \ref{sec:Application} and \ref{sec:Application2}, this technique may prove particularly valuable for the systematic construction of rotating non-BPS horizonless geometries in supergravity, which have attracted significant recent interest in the context of black hole microstructure. Such solutions are especially relevant for understanding coherent black hole microstates and probing horizon-scale quantum structure. To date, several classes of rotating horizonless solutions have been constructed, yet all known examples lie outside the black-hole cosmic censorship bound, either because of excessive angular momentum, as in the JMaRT solutions \cite{Jejjala:2005yu}, or because of excessive charges, as in the topological star \cite{Heidmann:2025pbb} and running bolt geometries \cite{Bena:2009qv,Bena:2025usu}. Consequently, these solutions cannot correspond to genuine microstates of astrophysical black holes. On the other hand, several classes of static horizonless geometries lying within the cosmic censorship bound have already been constructed \cite{Bah:2020ogh,Bah:2020pdz,Bah:2021owp,Bah:2021rki,Chakraborty:2025ger,Dima:2025tjz,Bah:2022yji,Bah:2023ows}. Our rotation-generating framework can be readily applied to systematically investigate the effect of angular momentum on these geometries and determine whether rotating smooth solutions within the black-hole regime can be achieved.

For simplicity, we restricted our analysis to five-dimensional minimal supergravity. However, many other compactifications of string theory and M-theory admit nonlinear sigma-model formulations with sufficiently large hidden symmetry structures. It would therefore be interesting to investigate how our construction generalizes to broader classes of supergravity theories and asymptotic geometries. In particular, this may allow one to extend the method beyond $\mathbb{R}^{1,3}$ and $\mathbb{R}^{1,4}$. From a holographic perspective, an especially interesting direction would be to investigate whether similar techniques can be developed for asymptotically $\mathbb{R}^{1,9}$ geometries whose decoupling limit yields black holes in AdS$_5\times$S$^5$ carrying angular momentum along the S$^5$.

It is also interesting to clarify the relation between the present construction and the class of TsT-type solution-generating transformations in string theory. In the standard TsT procedure, one performs a T-duality transformation, applies a coordinate shift along a second commuting isometry, and dualizes back; this provides a simple and powerful way of generating inequivalent string backgrounds, most famously the Lunin-Maldacena deformation and its O(2,2) formulation \cite{Lunin:2005jy,Frolov:2005dj,Catal-Ozer:2005dux}. Our construction exhibits a closely related conjugation structure: a Harrison-like transformation first maps the solution to a different asymptotic frame, a coordinate shift is then performed in that frame, and the inverse transformation returns the solution to the original asymptotic class. The essential difference is that the transformations employed here are not ordinary T-dualities acting on a two-torus, but hidden-symmetry transformations of the dimensionally reduced supergravity theory involving the Ehlers-Harrison sector. In this sense, the present map may be viewed as an analogue of TsT, with the O(d,d) duality group replaced by the whole sigma-model symmetry group. This perspective suggests several interesting questions. One could investigate the precise group-theoretic interpretation of the rotation-generating map, compare it with the $O(2,2)$ description of TsT and its relation to abelian Yang-Baxter deformations \cite{Osten:2016dvf}, and explore whether mixed constructions combining TsT and our transformations generate larger families of rotating solutions. Such connections may ultimately provide a broader framework for understanding how hidden symmetries, dualities, and coordinate transformations combine to generate nontrivial solutions in supergravity.

Finally, it is worth emphasizing that the systematic and algebraic nature of our construction makes it particularly well suited for modern computational and AI-assisted approaches to theoretical physics. Recent years have witnessed rapid progress in the use of AI tools for advanced research problems, and our framework provides a natural setting in which such methods could be explored. Indeed, the procedure involves many technical but highly structured steps and can be applied to a vast landscape of seed geometries. This opens the possibility of using AI-assisted techniques to efficiently scan large classes of solutions, explore their physical properties, and identify new rotating geometries that may otherwise remain difficult to construct analytically. In this sense, our work not only introduces a new solution-generating framework, but also provides a promising arena for combining hidden-symmetry methods in gravity with emerging AI-based approaches to theoretical physics.

\acknowledgments
SC, PH, and GP are supported by the Department of Physics at The Ohio State University. SC would like to acknowledge
the support provided by Anusandhan National Research Foundation (ANRF), India, through the Ramanujan fellowship grant RJF/2025/000037. SC is also supported by the Department of Physics at the Indian Institute of Technology, Kharagpur.

\appendix

\section{Sigma model from minimal five-dimensional supergravity}\label{app:5d}

In this appendix, we provide additional details on the sigma model arising from five-dimensional minimal supergravity upon reduction to three dimensions.

\subsection{Matrix representation}

Reducing five-dimensional minimal supergravity to three dimensions yields a nonlinear sigma model with action
\begin{equation}
    \mathcal{S}_3=\frac{1}{16 \pi G_3} \int\left(R_3 \star 1+\frac{1}{8} \operatorname{Tr}\left[d \mathcal{M}^{-1} \wedge \star d \mathcal{M}\right]\right) .  \label{app:SigmaModAction}
\end{equation}
The coset representative $\mathcal{M}$ depends on the eight scalar fields
$(Z_0, Z, \mu, A_t, A_\psi,\Omega_t,\Omega_\psi,a)$ introduced in Section \ref{eq:ReviewSM}. It can be written in terms of two $4\times4$ block matrices, $P=P^T$ and $Q^T=-Q$, as
\begin{equation} \label{app:MatrixM}
\mathcal{M} \equiv\left(\begin{array}{cc}
P & P Q \\
-Q P & P^{-1}-Q P Q
\end{array}\right), \quad \mathcal{M}^{-1}=\left(\begin{array}{cc}
P^{-1}-Q P Q & -Q P \\
P Q & P
\end{array}\right).
\end{equation}
The matrices $P$ and $Q$ are given by
\begin{equation}
\begin{aligned}
Q  \equiv\left(\begin{array}{cccc}
0 & \widetilde{A}_\psi & A_t & -B \\
* & 0 & B & -\Omega_\psi+\widetilde{A}_\psi B\\
* & * & 0 & -\Omega_t+A_t B\\
* & * & * & 0
\end{array}\right), \quad 
P  \equiv\left(\begin{array}{cc}
1-\Phi^T \Lambda \Phi & \Phi^T \Lambda \Psi \\
\Psi^T \Lambda \Phi & -\Psi^T \Lambda \Psi
\end{array}\right),
\end{aligned}   
\end{equation}
where we have introduced $\widetilde{A}_\psi \equiv A_\psi+\mu A_t$, together with
\begin{equation}
    \Psi=\left(\begin{array}{ccc}
    -1 & 0 & 0 \\
    0 & 1 & 0 \\
    A_t & -\widetilde{A}_\psi & -1
    \end{array}\right), \quad \Phi=\left(\begin{array}{c}
    -\widetilde{A}_\psi \\
    A_t \\
    B
    \end{array}\right), \quad \Lambda=Z_0 Z\left(\begin{array}{cc}
     h & 0 \\
    0 & 1
    \end{array}\right) .
\end{equation}
One can verify that $\mathcal{M}$ indeed belongs to the coset associated with $G_{2(2)}$, since it satisfies
\begin{equation}
    \mathcal{M}^{-1}=\left(\begin{array}{cc}
    0 & \mathbb{I}_4 \\
    \mathbb{I}_4 & 0
    \end{array}\right) \mathcal{M}^T\left(\begin{array}{cc}
    0 & \mathbb{I}_4 \\
    \mathbb{I}_4 & 0
    \end{array}\right), \quad \operatorname{det} \mathcal{M}=1~,
\end{equation}
together with the symmetry condition, $\mathcal{M}^T=\mathcal{M}$.

The equations of motion of five-dimensional supergravity take the compact form
\begin{equation}
    d\left(\mathcal{M}^{-1} \star d \mathcal{M}\right)=0, \qquad\left(R_3\right)_{\mu \nu}=-\frac{1}{8} \operatorname{Tr}\left[\partial_\mu \mathcal{M}^{-1} \partial_\nu \mathcal{M}\right]  .
\end{equation}
The first equation implies the existence of a dual matrix-valued one-form $\mathcal{N}$, which encodes the one-form fields of the five-dimensional solution, namely $(\omega_t,\omega_\psi,a)$:
\begin{equation} \label{app:MatrixN}
d \mathcal{N} \equiv \mathcal{M}^{-1} \star d \mathcal{M}.
\end{equation}

\subsection{Recovering the supergravity fields}

While the previous subsection explains how to construct the matrices $\mathcal{M}$ and $\mathcal{N}$ from the supergravity fields, it is equally important to extract the fields from a given coset representative. Using the results of \cite{Chakraborty:2025ger}, one finds
\begin{align}
Z_{0}&=\sqrt{\mathcal{M}_{42}^{2}-\mathcal{M}_{44}\mathcal{M}_{22}},\quad Z=-\frac{\mathcal{M}_{44}}{\sqrt{\mathcal{M}_{42}^{2}-\mathcal{M}_{44}\mathcal{M}_{22}}},\quad\mu=\frac{\mathcal{M}_{32}\mathcal{M}_{44}-\mathcal{M}_{42}\mathcal{M}_{43}}{\mathcal{M}_{42}^{2}-\mathcal{M}_{44}\mathcal{M}_{22}},\nonumber \\A_{t}&=-\frac{\mathcal{M}_{42}}{\mathcal{M}_{44}},\quad A_{\psi}=\frac{\mathcal{M}_{44}\mathcal{M}_{21}-\mathcal{M}_{41}\mathcal{M}_{42}}{\mathcal{M}_{42}^{2}-\mathcal{M}_{44}\mathcal{M}_{22}},\quad \Omega_t=\frac{\mathcal{M}_{44}\mathcal{M}_{47}-\mathcal{M}_{41}\mathcal{M}_{42}}{\mathcal{M}_{44}^2},\\
\Omega_\psi&= B\left(\frac{\mathcal{M}_{44}\mathcal{M}_{21}-\mathcal{M}_{41}\mathcal{M}_{42}}{\mathcal{M}_{24}^2-\mathcal{M}_{44}\mathcal{M}_{22}}+A_t\frac{\mathcal{M}_{32}\mathcal{M}_{44}-\mathcal{M}_{42}\mathcal{M}_{34}}{\mathcal{M}_{24}^2-\mathcal{M}_{44}\mathcal{M}_{22}}\right)+\frac{\mathcal{M}_{46}}{\mathcal{M}_{44}},\quad B=\frac{\mathcal{M}_{41}}{\mathcal{M}_{44}}, \nn \\
\omega_t&=-\mathcal{N}_{74}, \quad \omega_\psi=-\mathcal{N}_{64}, \quad a=-\mathcal{N}_{14}.\nn
\end{align}

\subsection{Generators of $G_{2(2)}$}
\label{App:GenerExpl}

The action \eqref{app:SigmaModAction} is invariant under global $G_{2(2)}$ transformations. Starting from a seed solution characterized by the matrices $(\mathcal{M}_0,\mathcal{N}_0)$, a new solution is generated through
\begin{equation}
    \mathcal{M}=g^T \mathcal{M}_0 g, \quad \mathcal{N}=g^{-1} \mathcal{N}_0 g, \quad g \in G_{2(2)}, \quad d s_3^2=d s_{30}^2 ,
\end{equation}
where $d s_3^2$ and $d s_{30}^2$ are the three-dimensional base that remain invariant.

The group elements are obtained by exponentiating the Lie algebra $\mathfrak{g}_{2(2)}$, which is a 14-dimensional algebra generated by the $8\times8$ matrices
\begin{equation}
    \mathcal{T}=\left\{\mathcal{H}_\pm, \mathcal{P}_{ \pm}, \mathcal{W}_{ \pm}, \mathcal{Z}_{ \pm}, \mathcal{O}_{ \pm i}, \mathcal{X}_{ \pm}\right\}~,
\end{equation}
with $i=1,2$. A generic group element is then written as
\begin{equation}
    g=e^{\mathfrak{g}} \in G_{2(2)}, \quad \text { where } \quad \mathfrak{g} \in \mathfrak{g}_{2(2)}~.
\end{equation}
The generators $\mathcal{H}_\pm$ span the Cartan subalgebra, while
$\mathcal{P}_{+}, \mathcal{W}_{+}, \mathcal{Z}_{+}, \mathcal{O}_{+i}, \mathcal{X}_{+}$
and
$\mathcal{P}_{-}, \mathcal{W}_{-}, \mathcal{Z}_{-}$, $\mathcal{O}_{-i}, \mathcal{X}_{-}$
correspond respectively to the positive and negative roots. The latter are related by matrix transposition,

To construct the generators explicitly, we first introduce the elementary matrices $E_{ij}$, with $i,j=1,\dots,8$, defined by
\begin{equation}
    \left(E_{i j}\right)_{k l}=\delta_{i k} \delta_{j l}, \quad k, l=1,2, \cdots, 8 .
\end{equation}
The generators of $\mathfrak{g}_{2(2)}$ can then be written compactly as follows:
\begin{itemize}
    \item The Cartan generators:
    \begin{equation}
    \mathcal{H}_+ \equiv 
    E_{44}
    -E_{88}
    +E_{33}
    -E_{77}, 
    \quad 
    \mathcal{H}_{-} \equiv 
    E_{44}
    -E_{88}
    +
    E_{22}-E_{66}  ~.   
    \end{equation}
    \item The $\mathcal{P}$-generators:
    \begin{equation}
    \begin{array}{lll}
    \mathcal{P}_{+} \equiv E_{45}-E_{18}+ E_{41}-E_{58}+ E_{27}-E_{36}, \\
    \mathcal{P}_{-} \equiv E_{54}-E_{81}+  E_{14}-E_{85} +  E_{72}-E_{63} .
    \end{array}
    \end{equation}
    \item The $\mathcal{W}$-generators:
    \begin{equation}
    \begin{aligned}
    & \mathcal{W}_{+} \equiv E_{56}-E_{21}+ E_{16}-E_{25}+ E_{43}-E_{78}, \\
    & \mathcal{W}_{-} \equiv E_{65}-E_{12}+  E_{61}-E_{52}+E_{34}-E_{87}.
    \end{aligned}
    \end{equation}
    \item The $\mathcal{Z}$-generators:
    \begin{equation}
    \begin{array}{lll}
    \mathcal{Z}_{+} \equiv E_{57}-E_{31}+E_{17}-E_{35}+ E_{68}-E_{42}, \\
    \mathcal{Z}_{-} \equiv E_{75}-E_{13}+ E_{71}-E_{53}+ E_{86}-E_{24} .
    \end{array}
    \end{equation}
    \item The $\mathcal{O}$-generators:
    \begin{equation}
    \mathcal{O}_{+1} \equiv E_{46}-E_{28}, \quad \mathcal{O}_{+2} \equiv E_{47}-E_{38}, \quad \mathcal{O}_{-1} \equiv E_{64}-E_{82}, \quad \mathcal{O}_{-2} \equiv E_{74}-E_{83} .
    \end{equation}
    \item The $\mathcal{X}$-generators:
    \begin{equation}
    \mathcal{X}_{+} \equiv E_{67}-E_{32}, \quad \mathcal{X}_{-} \equiv E_{76}-E_{23} .
    \end{equation}
\end{itemize}

\section{Coset matrices for various seed solutions}
In this appendix, we list the coset matrices $\mathcal{M}$ and $\mathcal{N}$ corresponding to the three seed geometries considered in this paper.

\paragraph{Schwarzschild.}

For the Schwarzschild geometry trivially embedded in five dimensions \eqref{eq:Schwarzschild4d}, the matrices take the form
\begin{align}
    \mathcal{M}
    &=
    \text{diag}
    \left(
    1,1,-\frac{1}{f},-\frac{1}{f},1,1,-f,-f
    \right) \, , \qquad \mathcal{N}
    &=\text{diag}
    \left(
    0
    ,
    0
    , 
    h
    ,
    h
    ,
    0
    ,
    0
    ,
    -h,-h\right) \, . \label{eq:SchwarzschildMandN}
\end{align}
with $f = 1-\frac{2m}{r}$ and $h =  2m\cos{\theta}d\phi$.

\paragraph{Schwarzschild-Tangherlini.} 

For the static five-dimensional black hole \eqref{eq:SchwarzschildTan}, written in Hopf-fibered coordinates, one finds
\begin{align}
    \mathcal{M}
    &=
    \text{diag}
    \left(
    1, 
    \frac{4}{r^2},
    -\frac{1}{f},
    -\frac{4}{r^2f},
    1,
    -\frac{m}{4f},
    -f,
    \frac{m}{4}
    \right)
    +
    \frac{1}{f}(E_{46}+E_{64})
    +
    (E_{28}+E_{82}) \, , \nn
    \\
    \mathcal{N}
    &=
    \text{diag}
    \left(
    0, 
    -h,
    h,
    0,
    0,
    h,
    -h,
    -0
    \right)
    +\frac{4h}{m}\, (E_{82}-E_{64})\, , \label{eq:SchwarzschildTMandN}
\end{align}
with $f = 1-\frac{m}{r^2}$ and $h = \frac{1}{4}m \cos{\theta}d\phi$. 

\paragraph{Weyl solutions.}

Finally, for generic four-dimensional Weyl solutions trivially embedded in five dimensions \eqref{eq:WeylMetric}, the matrices are
\begin{align}
    \mathcal{M}
    &=
    \text{diag}
    \left(
    1,1,-\frac{1}{U},-\frac{1}{U},1,1,-U,-U
    \right) \, , \nn
    \\
    \mathcal{N}
    &=\text{diag}
    \left(
    0
    ,
    0
    , 
    -Vd\phi
    ,
    -Vd\phi
    ,
    0
    ,
    0
    ,
    Vd\phi,Vd\phi\right) \, . \label{eq:SchwarzschildMandN}
\end{align}
Here, $U$ denotes the harmonic function satisfying the Laplace equation \eqref{eq:WeylEq}, while $V$ is its dual potential defined in \eqref{eq:WeylDual}.

\section{Rotating Weyl solution in five dimensions}
\label{app:RotWeyl}

In this section, we provide the explicit form of the five-dimensional solutions obtained by applying the rotation-generating technique to the Weyl solution introduced in Section \ref{sec:RotWeyl}. The metric and the U(1) gauge field take the form
\begin{align} 
ds_{5}^{2}&=\left(d\psi-\mathcal{A}\right)^{2}+ds_{4}^{2},\qquad 
A=A_{t} \left(dt+\omega_{t}\right)+a,
\end{align}
where the four-dimensional fields, $ds_4^2$ and $\mathcal{A}$, are given in \eqref{eq:RotWeyl}. The remaining components are
\begin{align} A_{t}&=\frac{\Omega^{2}\sqrt{U}\left(T\bar{U}X+2\sqrt{U}S(V\bar{U}-2S)\right)}{4\Omega^{2}US^{2}+X^{2}},\\ a&=-\Omega\left(\frac{\Omega^{2}\left(2T^{2}+\rho^{2}\bar{U}^{2}(V\bar{U}-2S)(V\bar{U}-4S)\right)-2\rho^{2}U^{-1/2}\bar{U}+(V+2z)V}{2\Delta}+W\right). \nn \end{align}
One can verify that the U(1) gauge field is dual to the Kaluza--Klein gauge field in four dimensions,
\begin{align}
\star_4 dA = d\mathcal{A},
\end{align}
so that, upon reduction to four dimensions, the two fields can be combined into a single gauge field as presented in Section \ref{sec:RotWeyl}.

\bibliography{RotBH}
\bibliographystyle{JHEP}

\end{document}